\documentclass[conference]{IEEEtran}
\IEEEoverridecommandlockouts
\usepackage{cite}
\usepackage{amsmath,amssymb,amsfonts}
\usepackage{algorithmic}
\usepackage{graphicx}
\usepackage{textcomp}
\usepackage{xcolor}
\usepackage{cite}
\usepackage{mathtools}
\usepackage{makecell}
\usepackage{mathrsfs}

\def\BibTeX{{\rm B\kern-.05em{\sc i\kern-.025em b}\kern-.08em
    T\kern-.1667em\lower.7ex\hbox{E}\kern-.125emX}}

\begin{document}

\makeatletter
\newcommand{\linebreakand}{%
  \end{@IEEEauthorhalign}
  \hfill\mbox{}\par
  \mbox{}\hfill\begin{@IEEEauthorhalign}
}
\makeatother

\title{Use of Deep Neural Networks for Uncertain Stress Functions with Extensions to Impact Mechanics
}

\author{
\IEEEauthorblockN{Garrett Blum}
\IEEEauthorblockA{\textit{Mechanical Engineering} \\
\textit{Northwestern University}\\
Evanston, USA \\
garrettblum2024@u.northwestern.edu}
*Corresponding author
~\\
\and
\IEEEauthorblockN{Ryan Doris}
\IEEEauthorblockA{\textit{Product Development} \\
\textit{Team Wendy, LLC}\\
Cleveland, USA \\
rdoris@teamwendy.com}
% ~\\
\and
\IEEEauthorblockN{Diego Klabjan}
\IEEEauthorblockA{\textit{Industrial Engineering} \\
\textit{Northwestern University}\\
Evanston, USA \\
d-klabjan@northwestern.edu}
% ~\\
\linebreakand
\IEEEauthorblockN{Horacio Espinosa}
\IEEEauthorblockA{\textit{Mechanical Engineering} \\
\textit{Northwestern University}\\
Evanston, USA \\
espinosa@northwestern.edu}
% ~\\
\and
\IEEEauthorblockN{Ron Szalkowski}
\IEEEauthorblockA{\textit{Product Development} \\
\textit{Team Wendy, LLC}\\
Cleveland, USA \\
rszalkowski@teamwendy.com}
}

\maketitle

\begin{abstract}
Stress-strain curves, or more generally, stress functions, are an extremely important characterization of a material’s mechanical properties. However, stress functions are often difficult to derive and are narrowly tailored to a specific material. Further, large deformations, high strain-rates, temperature sensitivity, and effect of material parameters compound modeling challenges. We propose a generalized deep neural network approach to model stress as a state function with quantile regression to capture uncertainty. We extend these models to uniaxial impact mechanics using stochastic differential equations to demonstrate a use case and provide a framework for implementing this uncertainty-aware stress function. We provide experiments benchmarking our approach against leading constitutive, machine learning, and transfer learning approaches to stress and impact mechanics modeling on publicly available and newly presented data sets. We also provide a framework to optimize material parameters given multiple competing impact scenarios.  
\end{abstract}


\section{Introduction}

Stress functions, sometimes referred to as stress-strain curves, are a key tool to study the response of materials under loading regardless of application and serve as the base for finite element modeling. Such functions relate stress, the force per area, to strain, the percent deformation of the material in compression, tension, and torsion. Together, stress tensors completely describe the state of stress at a point inside a material in the deformed state, placement, or configuration. Alternatively, individual stress components can be examined in simplified scenarios as well such as in uniaxial tension or compression. Additional complexity exists when the effect of large deformations, high strain-rates, material parameters, or temperature on stress needs to be understood \cite{mechanics-of-materials}.

Historically, classical constitutive modeling has been the foundation for modeling stress functions. This classical approach is inherently parametric in formulation and is highly limited in scope and versatility. Suppose, for example, one is interested in characterizing a new material for which there does not exist a classical constitutive model. The work to adapt an existing approach to this new material would either be non-trivial or more likely require an entirely new model, an extensive process of derivation and validation. We are unaware of existing constitutive theory that generalizes across materials with the exception of relationships such as Hooke's law that are highly limited in domain (strain in the linear elastic region at constant strain-rate). We contend that such limited scopes are an issue with the classical constitutive theory.

In light of this issue, recent work has examined several attempts at incorporating ``big data" and machine learning to address these concerns. Ghaboussi et al. \cite{ghaboussi} suggested that the behavior of materials may be reflected by neural networks as early as 1991. Later, Al-Haik et al. \cite{ALHAIK2006} developed a neural network based model to predict the stress relaxation of polymer matrix composites, proving viability of the concept and demonstrating that neural networks can learn stress relationships without prior knowledge or assumptions of the functional form. Despite this, limited work has been completed to advance such approaches, especially for general use models that are applicable to wide varieties of materials.  

Moreover, both the classical constitutive models and machine learning approaches suffer from a third shortcoming: they are all point-estimates of stress. Material behavior is known to be uncertain, both spatially and point-wise \cite{Jeremić}. Such uncertainty is vital to understand particularly in design and optimization scenarios. Modeling the material's uncertainty allows for design problems to be redefined from optimizing the expected response of a material to optimizing the distribution of outcomes, including the quantification of confidence intervals. Moreover, such an understanding allows researchers to penalize high variability should it be undesirable in the application. This understanding is particularly important in applications such as helmets where small differences in kinematic response can lead to large differences in user safety (concussion vs traumatic brain injury vs death).

We address the three aforementioned issues: the lack of generalization offered by classical constitutive theory, the lack of generalized machine learning formulations, and the lack of material models that specifically account for the observed variation in material response. We present a deep neural network based approach that aims to solve the three concerns while providing versatility across fields as well as flexibility in defining both the domain and codomain. Additionally, we employ quantile regression to capture observed variance in the experimental data as a function of the parameter space.

Our method is widely applicable to the fields of mechanics of materials and material science, but we focus on impact mechanics as a particularly difficult proving ground requiring complex stress models with large deformations, strain-rates, and temperature ranges. Within impact mechanics, the kinematics and especially the acceleration of the impacting body are commonly studied due to their known correlations with safety particularly in car crash-worthiness and helmet safety \cite{impact-kin}. We provide a robust and fast numerical method for calculating impact kinematics not reliant on FEM or any specialized software derived from Newtonian mechanics. 

Our main contributions are summarized as follows.
\begin{enumerate}
    \item We contribute a deep quantile approach to model stress as a function of strain, strain-rate, temperature, and material parameters.
    \item We provide a formulation that captures variation in observed data without loss of accuracy for point estimation.
    \item A Newtonian mechanics approach to predict uniaxial impact kinematics from uncertain stress functions and initial conditions using stochastic differential equations is developed.
\end{enumerate}

In Section \ref{related_work}, we present a summary of existing models while demonstrating the lack of uncertainty based approaches. We then highlight the improvements that our uncertainty aware deep quantile regression approach provides. In Section \ref{approach_and_methods}, we introduce the deep learning formulation of the stress model and derive a theoretical extension to uniaxial impact mechanics using nonlinear stochastic differential equations. In Section \ref{experiments}, we provide details for three data sets on which we test and benchmark our approach before providing implementation details for our experiments. We then present our results. In Section \ref{discussion}, we compare our results with their respective benchmarks and demonstrate versatility and use-case through multi-objective optimization of material parameters under competing impact conditions on the the 98A data set.

\section{Related Work}\label{related_work}
Traditional constitutive models can be grouped into two categories: physics models and phenomenological models. First, the so-called physics models aim to describe the macroscopic behavior of materials based on the microstructures where all parameters in the models have physical meanings. For example, in glassy thermoplastics and rubbery materials respectively, most of the physical constitutive models are derived from the well known theories of Haward and Thackray \cite{haward-and-thackray} and Edwards and Vilgis \cite{EDWARDS1986}. Alternatively, phenomenological models aim to describe the experimental phenomenon by using mathematical formulations with or without some concepts of physical constitutive models. This approach includes many elasto-plsatic, viscoelastic, viscoplasic, hyperelastic, damage, and machine learning based approaches \cite{LingWuMei}. For example, the famous Johnson-Cook model for metals, as well as subsequent derivations, are phenomenological models \cite{johnson-cook}. As previously discussed, classical constitutive models are parametric in nature and often make assumptions regarding the form of the stress function. Our neural network based approach maintains the same input-output relationship while making no such assumptions which may offer significant benefits through applicability across various domains and materials.

As a representative constitutive model, we focus on the recent work of Johnson et al. \cite{JOHNSEN2019}, a leading thermo-elasto-viscoplastic constitutive model for polymers which employs 14 parameters. This approach makes assumptions about material response through its use of a hyper-plastic Hencky spring in series with two Ree-Eyring dashpots in addition to an eight-chain model to explain the stress response. This model is only applicable to polymer response and makes no mention of how the effect of material parameters should be taken into account. We choose Johnson's approach as a benchmark to represent a leading classical constitutive model. 

Particularly in recent years, machine learning models have been proposed as an alternative to traditional modeling techniques. First we note that convolutional neural networks and recurrent neural networks have been employed to examine spacially dependent and path dependent stress functions, respectively \cite{STOFFEL2020, GORJI2020}. However, the majority of the past studies do not include spacially dependent and path dependent stress functions which is the practice we follow. Within this scope, Li et al. \cite{LI2019320} propose a hybrid approach extending the Johnson-Cook model where a neural network is employed to capture unconventional effects of the strain-rate and temperature on the hardening response for DP steel. Similarly, Jordan et al. \cite{JORDAN2020} model stress in two parts: elastic deformation employes a temperature-dependent Hooke’s law, while the viscoelastic deformation is reflected by a neural network model. First, they mix in the Hooke’s law component rather than only using a neural network. Second, their formulation applies only to polypropylene. Instead, we propose a method that models stress entirely with a neural network and whose formulation does not constrain it to a single material or class of materials. Using their data set, we benchmark against this approach, particularly due to its wide acceptance and comparison with the aforementioned Johnson model.

As we focus on impact mechanics, we must note an alternative approach to achieving impact related measures that circumvent material modeling. Gongora et al. \cite{GONGORA2022} recently proposed a ``transfer learning" approach to directly predict peak acceleration of an impact for resin-based lattice structures. This approach takes select material parameters and a quasi-static stress-strain curve as inputs, performs principle component analysis (PCA), and applies a forward stepwise linear regression to predict peak acceleration. However, their approach is only capable of predicting impact related criteria for samples at a single impact velocity, drop mass, thickness, and area dictated by their experimental data. Extensions to vary the aforementioned parameters appear to be either nontrivial or require data sets orders of magnitude larger to observe the effects experimentally. Additionally, requiring an experimental stress-strain curve as input is highly limiting. As we model the stress function and use differential equations to yield impact kinematics, we eliminate all of the aforementioned concerns at the expense of more computation during training. Using their publicly available data set, we benchmark against this approach as well.

Finally, we mention that literature related to the uncertainty of the stress response is limited. The majority of studies aim to capture the uncertainty of estimated parameters of existing parametric functions for stress \cite{Jeremić, Ivan, BREWICK2018}. Additionally, several studies aim to capture and quantify measurement uncertainty from the experimental methodology \cite{Brizard}. Finally, several studies examine the propagation and significance of parameter uncertainty, emphasizing significant importance in capturing this variance \cite{Romero, Brizard}. However, all approaches make parametric assumptions for the stress function, in turn highly limiting the scope. We eliminate this constraint through our generalized formulation. We work under the \textit{a priori} assumption that there exists some variance in the stress response that is a function of the input parameters that we wish to model. To the best of our knowledge, this approach is the first of its kind.

\section{Modeling and Theory} \label{approach_and_methods}

We present two variants of a deep learning methodology to estimate stress as a state function, one using point estimation and the other employing quantile regression to capture uncertainty. Additionally, we present theory to extend such models to impact mechanics. 

\subsection{Stress as a Deep Neural Network}

We propose a generalized methodology for modeling state stress functions of the form  $\sigma(\epsilon, \dot{\epsilon}, T, \vec{p})$ using fully connected, feed-forward Deep Neural Networks (DNNs). 

Let $f_{DNN}$ be a DNN that takes a feature vector, $\vec{x}$, as input and outputs stress, $\vec{y}$. The optimal weights of the network are found through supervised learning by minimizing a loss function from data where both $\vec{x}$ and $\vec{y}$ are observed. 

We choose $\vec{x}$ to have the form 
$
    \vec{x} = \begin{bmatrix} \epsilon, & \dot{\epsilon}, & T, & \vec{p} \end{bmatrix} 
$
where $\epsilon \in [0,1]$ is strain in fractional percent, $\dot{\epsilon} \in {\rm I\!R}$ is strain-rate in $sec^{-1}$, $T \in {\rm I\!R}$ is temperature in degrees Celsius, and $\vec{p} \in {\rm I\!R}^{k}$ corresponds to $k$ material properties. The output, $\vec{y}$, has two forms. When we are interested in point-estimates of the stress function, $\vec{y} = [\sigma]$. When we are interested in estimating stress as well as its uncertainty, the output vector has the form
$
    \vec{y} = \begin{bmatrix} q_{1}, & ..., & q_{w} \end{bmatrix} \in {\rm I\!R}^{w}
$
which corresponds to the $w$ quantile estimates for stress, $q$, each with a given significance level, $\alpha$. 

We employ LDS, or Label Distribution Smoothing, a methodology developed to combat label imbalance in DNNs \cite{yang2021delving}. A Gaussian kernel with 100 buckets is used where the kernel size and the standard deviation are hyperparameters.

\subsection{Impact Mechanics Theory}

When a material is deformed under impact, the material's stress function plays a key role in defining the kinematics of said impact. When studying impact mechanics acceleration, $a$, is needed. Here, we propose how to calculate it from the stress function, $\sigma$, or its approximation. We consider a sample of a given, deformable material with known thickness, $d$, area, $A$, and compressive stress function, $\sigma(\epsilon, \dot{\epsilon}, T, \vec{p})$, resting on a fixed, undeformable, flat surface as shown in Figure \ref{impact-mech-theory-schematic}. An undeformable impacting anvil with a flat bottom surface of known mass, $m$, and initial velocity, $v_0$, exists directly above the sample. Without energy loss, said anvil is uniaxially constrained to move only in the vertical direction. The entire system acts under the force of gravity with gravitational constant, $g=9.81 m/s^2$. This formulation is based on the uniaxial impact tests later discussed and shown in Figure \ref{experimental_setup_fig}. We derive an equation for the acceleration of the anvil as a function of time, $a(t)$, throughout the duration of the impact.  

\begin{figure}[htbp]
\centering
\includegraphics[width=0.75\columnwidth]{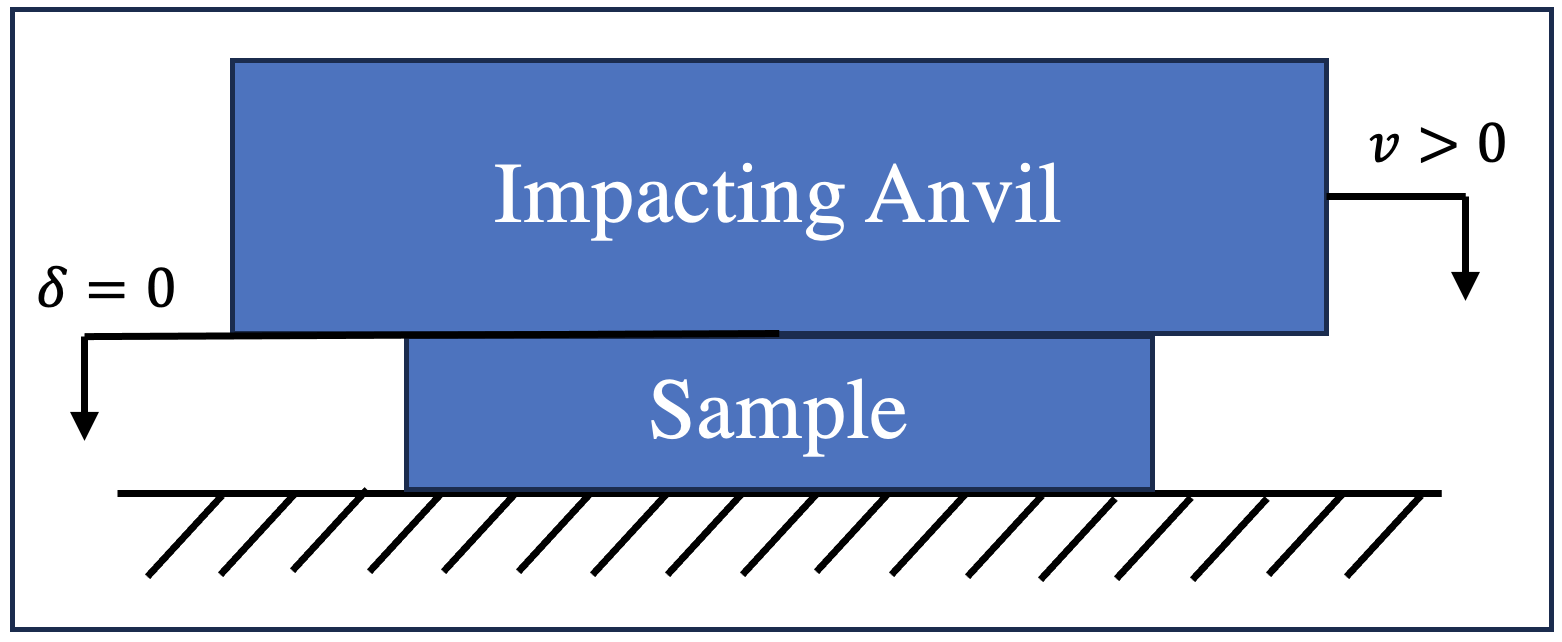}
\caption{Schematic to illustrate uniaxial impact mechanics derivation.}
\label{impact-mech-theory-schematic}
\end{figure}

We begin by making five simplifying assumptions. First, we disregard stress wave propagation within the deformable sample. Second, we disregard the edge effects of the deformable sample. Third, we disregard the mass and by extension the momentum of the deformable sample. Finally, we assume that the deformable sample has spacialy uniform material properties ($\vec{p}$) and that $T$ is constant throughout the duration of the impact. 

Approaching the problem using Newtonian mechanics, we consider the sum of the forces on the impacting anvil as

\begin{equation}
ma = F - mg
\label{sum-forces-eq}
\end{equation}
with F being the force of the sample on the anvil. Recalling that $A \cdot \sigma = F$, we substitute in terms of stress:

\begin{equation}
ma = A \cdot \sigma(\epsilon, \dot{\epsilon}, T, \vec{p}) - mg.
\label{sum-forces-eq-2}
\end{equation}

By continuity, both the positions and velocities of the bottom of the falling anvil and the top surface of the deformable sample are equal. We then write the strain and strain-rate of the sample, as functions of time, as follows:
\begin{gather}
    \label{strain(time)_eq}
    \epsilon(t) = \frac{\delta(t)}{d} \\
    \dot{\epsilon}(t) = \frac{v(t)}{d}
    \label{strain-rate(time)_eq}
\end{gather}
where $\delta$ is defined to be $0$ at the initial location of the top of the deformable sample, where positive is down, and we have $v=\frac{d\delta}{dt}$, $a=\frac{dv}{dt} = \frac{d^2\delta}{dt^2}$. Substituting into (\ref{sum-forces-eq-2}) with simplification yields (\ref{tbd2}), a nonlinear, second-order differential equation for $a(t)$ that accounts for strain-rate hardening and temperature sensitivity of the sample as well as material parameters. If the stress function, $\sigma$, accounts for uncertainty, then $a(t)$ becomes a stochastic differential equation (SDE). 

\begin{equation}
a(t) = \frac{A}{m} \cdot \sigma\left(\frac{\delta(t)}{d}, \frac{v(t)}{d}, T, \vec{p}\right) - g
\label{tbd2}
\end{equation}

Depending on one's choice for $\sigma$, there may exist an analytical solution to this differential equation. However, since we employ a neural network to model stress, we solve this equation numerically using an Explicit Euler scheme: 

\begin{gather}
  \label{EE1}
  \delta_{t+1} = \delta_{t} + \Delta t \cdot v_{t} \\
  \label{EE2}
  v_{t+1} = v_{t} + \Delta t \cdot a_{t} \\
  \label{EE3}
  a_{t+1} = \frac{A}{m} \cdot \sigma \left(\frac{\delta_{t+1}}{d}, \frac{v_{t+1}}{d}, T, \vec{p}\right) - g.
\end{gather}

In the case $\sigma$ is stochastic, we assume: 
\begin{equation}
    \sigma(\epsilon, \dot{\epsilon}, T, \vec{p}) = \mathcal{N}\left(\mu(\epsilon, \dot{\epsilon}, T, \vec{p}), \Sigma(\epsilon, \dot{\epsilon}, T, \vec{p})\right).
\end{equation}

We nominally estimate $\mu, \Sigma$ based on experiments using quantile estimates each of which is spaced at significance levels to be approximately one standard deviation apart:

\begin{gather}
    \mu(\epsilon, \dot{\epsilon}, T, \vec{p}) \approx q_{\alpha = 0.5} \\
    \Sigma(\epsilon, \dot{\epsilon}, T, \vec{p}) \approx \left(\frac{1}{Q-1}\sum_{i=1}^{Q-1} (q_{i+1}-q_i) \right)^2.
\end{gather}

We run 1,000 simulations to solve for $\sigma$ where in each iteration we sample on $\mathcal{N}$ and solve (\ref{EE1})-(\ref{EE3}). We then average all of the computed $a(t)$'s. We emphasize that formulating the problem as a SDE enforces that each simulation maintains physical accuracy as the integral of $a(t)$ until densification is the impact velocity. All simulations are conducted at the same time step ($\Delta t$) as the experimental data for a given data set.

\section{Experiments} \label{experiments}

\subsection{Data}

We investigate the performance of our deep learning approach on three data sets to benchmark performance and compare with existing methodologies. 

First, we examine Jordan's data set of polypropylene samples in tension. By extension, we  benchmark against their neural network based approach as well as Johnson's recently-proposed thermos-elastic-viscoplastic constitutive model for polymers \cite{JORDAN2020, JOHNSEN2019}. The data set contains 40 tensile tests (5,757 data points) of samples with consistent material properties at stain-rates ranging between $~10^{-3}$ and $~10^{-1} sec^{-1}$ and temperatures between 25$^{\circ}$C and 80$^{\circ}$C. 

Second, we examine the Gongora's data set of a parametric family of lattices thus benchmarking against their transfer learning approach to directly predict peak acceleration during impacts \cite{GONGORA2022}. The publicly available data set contains 833 quasi-static compression tests at $~0.0018 sec^{-1}$ and 270 impact tests of which we include 193 (1,812,480 data points) for training. All impacts are conducted with an initial impact velocity of $4.3m/s$ and drop mass of $3.104kg$ with a capture rate of 2,000kHz. 

We only employ the dynamic drop test data. To calculate strain and strain-rate we employ (\ref{strain(time)_eq}) and (\ref{strain-rate(time)_eq}), respectively. To calculate stress from these acceleration data consistent with accepted methodologies, the following is used:

\begin{equation}
    \sigma(t) = \frac{m \cdot a(t)}{A} \label{stress-from-accel}.
\end{equation}
We emphasize that drop tower tests have previously been employed and are an accepted means to characterize a material at intermediate strain-rates \cite{Brown_Drop_Tower, ISLAM2020}.

Finally, we collect a new data set of a previously characterized open-celled polyurethane foam called 98A \cite{Mines_Charecterization}. The foam's properties are controlled by two independently determined variables, one having to do with its chemistry (``index") and the other having to do with its pore size (``altered pore size"). Additionally, a one hot encoded variable regarding when the samples are manufactured is included (three possible values).

\begin{figure}[htbp]
\centering
\includegraphics[width=0.6\columnwidth]{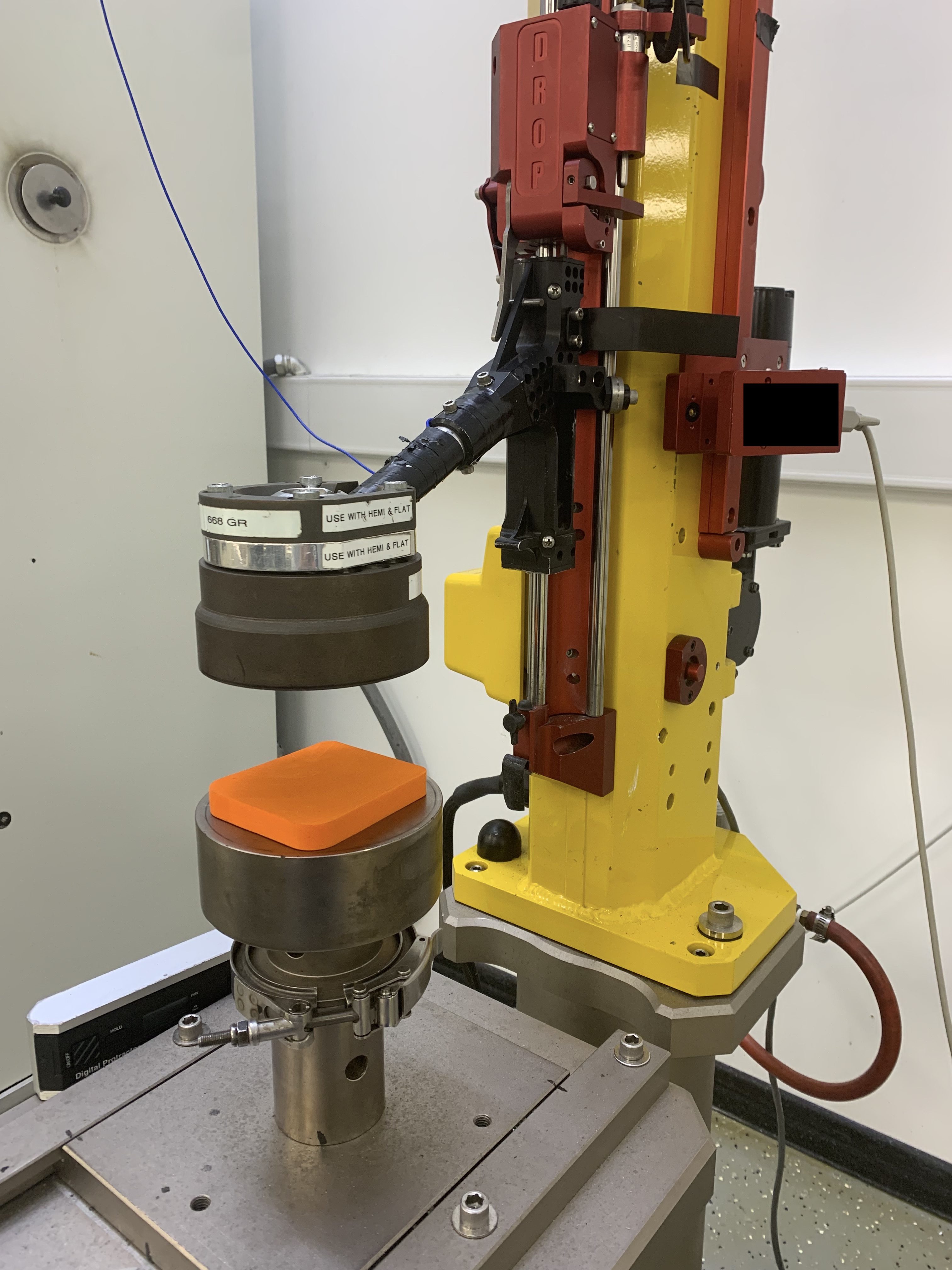}
\caption{Image of the Cadex drop tower used for experimental data collection for the 98A data set.}
\label{experimental_setup_fig}
\end{figure}

We conduct drop tests using a Cadex uniaxial monorail machine consistent with existing methodologies as shown in Figure \ref{experimental_setup_fig} \cite{Brown_Drop_Tower, ISLAM2020}. Samples of known thickness and area are placed on the fixed lower anvil. The drop assembly ($m=5kg$) is raised to a height corresponding to the desired impact velocity. Acceleration data is captured at 33kHz, filtered using a CFC 1000 per SAE J211, and the impact velocity is recorded using a timed laser gate \cite{SEA_standard}. Samples are preconditioned in three temperature environments: ambient is a nominal 21$^{\circ}$C, cold is a nominal -10$^{\circ}$C, and hot is a nominal 54$^{\circ}$C for a minimum of 12 hours. Care is taken to remove a sample from conditioning and test it quickly as to minimize heat transfer (typically under 20 seconds).

In total, 2,459 impacts are conducted: 2,193 at ambient, 132 at hot, and 134 at cold. This results in 528,097 datapoints. The range on index examined is $[79, 130]$ and altered pore size is on the interval $[560, 700]$. These represent the reasonable range of manufacturable inputs. Impact velocities range from $3.18m/s$ to $5.67m/s$ with peak accelerations ranging from $173.5G$ to $701.2G$. Similar to the Gongora lattice data set, we employ (\ref{strain(time)_eq}), (\ref{strain-rate(time)_eq}), and (\ref{stress-from-accel}) to calculate strain, strain-rate, and stress from these data.

\subsection{Implementation}

\subsubsection{Training and Metrics}

Cross-validation is employed to ensure generalization of the model where the data is split into training, validation, and testing subsets. We opt to define the test sets based on unique sets of $T$ and $\vec{p}$ unless otherwise noted. Doing so enforces that cross-validation measures how well the model can generalize to a set of material properties or temperatures not seen during training rather than to new values of strain or strain-rate for material properties and temperatures known during training. We believe this subtle distinction is more inline with such models' use cases. We note that we are unaware of other work that cross-validates using this technique but emphasize its benefits particularly when working with neural networks where over-fitting is especially a concern. Additionally, we examine these models as ensembles where the ensemble estimates are the simple element-wise averages of the individual folds. 

Mini-batch back-propagation with the Adam optimizer is used to train the DNNs. The batch sizes are 8, 32, and 16 for the Jordan data set, the Gongora data set, and the 98A data set, respectively. Mean absolute error (MAE or L1) loss is employed for point-estimation models and quantile (sometimes referred to as pinball) loss is employed for uncertainty aware networks. We examine $\alpha = \{0.0228, 0.1587, 0.5, 0.8413, 0.9772\}$ to nominally estimate one and two standard deviation confidence intervals. The networks are trained for 100, 50, and 100 epochs, respectively. 

Kubernetes is used for container management. All training for the Jordan and 98A data sets is conducted on pods with 64Gi of memory using Intel Core i7-7820X CPUs at 3.60GHz, 8 cores each. Additionally, we employ a single NVIDIA GeForce RTX 2080 Ti GPU (11,019MiB). For the Gongora data set, we employ the same hardware but with 48Gi of memory allocated. 

Consistent with other ``big data" and machine learning literature, we also report the root-mean-squared-error (RMSE), and $R^2$ along with the standard deviation of stress for point estimates \cite{2022arXiv221102989J}. For interval estimates, we report quantile loss and absolute coverage difference (ACD) \cite{MAKRIDAKIS2020}. For peak linear acceleration, we also report root-mean-squared-percentage-error (RMSPE) consistent with the domain's common reporting practice \cite{GONGORA2022}.

\subsubsection{Hyperparameters}

We employ a sequential grid search to optimize hyperparameters. For our formulation of $f_{DNN}$, we have five hyperparameters: the architecture, the learning rate ($\gamma$), the dropout rate, Lable Distribution Smoothing (LDS) kernel size, and $\sigma_{LDS}$. 

To expedite hyperparamter selection, we employ a randomly down-sampled subset within each CV fold for the Gongora data set (2.5\% of data) and the 98A data set (5\% of data). We do not downsample the Jordan data set due to the small number of samples. Additionally, we train the networks for fewer epochs during hyperparameter optimization than final models: 50, 25, and 50, respectively. 

First, a random search of 50 trials is employed holding the architecture constant at $(75, 50, 25)$ neurons per layer and the dropout rate at $0.25$. The learning rate is randomly selected from the logarithmic distribution on the interval $[10^{-6}, 0.1]$, LDS kernel size from the odd integers on the interval $[1, 15]$, and $\sigma_{LDS}$ from the uniform distribution on the interval $[1, 5]$. A 20-fold cross-validation is completed for each trial and hyperparameters with the best CV test loss are taken as the baseline.

Next, a sequential grid search is completed for all 5 hyperparameters starting with the architecture. Exactly 50 candidate architectures are selected ranging from 165 to 661,565 total trainable parameters with between 1 and 5 hidden layers. The architecture with the best CV test loss replaces the baseline in the aforementioned set. 

With the updated set of hyperparameters, the constant learning rate is next optimized. Again, 50 candidates are selected uniformly from the logarithmic distribution on the interval $[10^{-8}, 1.0]$. Similarly, the learning rate with the best CV test loss replaces the baseline in the aforementioned set.

This process of completing a grid search for a given hyperparameter while holding all others constant with the current best to update the baseline is repeated with the remaining three hyperparameters in order of the dropout rate, LDS kernel size, and $\sigma_{LDS}$. Here, 20 candidates are selected uniformly for dropout rate from the uniform distribution on the interval $[0, 0.95]$. Likewise, 20 candidates are selected uniformly for the LDS kernel size from the uniform distribution of odd integers on the interval $[1, 39]$. Finally, 20 candidates are selected uniformly for $\sigma_{LDS}$ from the logarithmic distribution on the interval $[0.5, 20]$. Hyperparameters are optimized independently for each data set.

\subsubsection{Multiobjective Optimization}

Often for impact applications, we wish to optimize a material's parameters given several competing impact conditions. Such conditions could include various temperatures, drop masses, and/or impact velocities. For the 98A data set, we present a brief optimization to demonstrate the use case of these models.

We frame this optimization based on helmet standards \cite{IHSP_spec} where an impact's safety is based on minimizing peak linear acceleration across a matrix of impact velocities and temperatures. Additionally, there are pass-fail thresholds for each impact velocity which remain constant across temperatures. We note such approaches are criticized measures of performance as they do not account for impact duration or rotational acceleration and the development of alternative measures remains an active area of research \cite{ESTRADA2021, background-metrics}. However, since pass-fail peak linear acceleration based criteria are prevalent in accepted testing standards, we formulate this optimization as such.

We examine 6 impact metrics comprised of the combinations of two impact velocities ($3.0m/s$ and $4.1m/s$) across three temperature conditions ($-10^{\circ}$C, $21^{\circ}$C, $54^{\circ}$C for cold, ambient, and hot, respectively). We maintain the sample thickness to be $25mm$, the area to be $7,000mm^2$, and the drop mass to be $5kg$. The acceptable threshold for $3.0m/s$ impacts is 120G and for $4.1m/s$ impacts it is 150G. We opt for a combined grid and random search to explore the search space. One could employ a stochastic optimization scheme if they are solely concerned with identifying the optimal material parameters rather than exploring the parameter space.

We define the smooth sigmoid-inspired loss function for a given impact metric to be
\begin{equation}
    l_{i,j}(x) = \frac{a}{1+e^{-b(x-T)}}
\end{equation}
where $x$ is the peak acceleration of the impact metric $i$ examined at quantile $j$, $T$ is the acceptable threshold for the impact, and $a,b$ are hyperparameters that control the maximum value and the slopes around the threshold value, respectively. We choose $a=100$ and $b=0.125$ for our experiments. These values give stable and useful results.

When concurrently optimizing for $S$ impact metrics ($S=6$ in our case) where we have $Q$ quantile measures of peak acceleration, we define the combined loss function to be the unweighted average as follows
\begin{equation}
\label{loss_func}
    \mathscr{L} = \frac{1}{SQ} \cdot \sum_{i=1}^{S} \alpha_{i} \sum_{j=1}^{Q}l_{i,j}(x)
\end{equation}
where $\alpha_i$ is the importance weight on impact metric $i$. In our experiments, we set $\alpha_i = 1$ for every $i$ since the IHPS standard \cite{IHSP_spec} gives equal importance to the metrics.

\subsection{Results}

We present results in three parts. First, we compare results between our deep quantile stress model against a state-of-the-art constitutive model and a hybrid constitutive-machine learning approach. Next, we compare against a transfer learning methodology that directly predicts peak acceleration, circumventing stress models and time-dependent kinematics. Finally, we present a third data set to demonstrate the full versatility of our approach including a demonstrative optimization. 

\subsubsection{Benchmarking against Jordan and Johnson Models}

The results of the hyperparameter optimization using the L1 loss function are shown in Figure \ref{jordan_hyperparams}. The chosen hyperparamters are as follows: the model architecture is $(100, 50)$, the learning rate is $2.154\cdot10^{-2}$, the dropout rate is $0.15$, the LDS kernel size is $3$, and $\sigma_{LDS}$ is $6.239$. These hyperparameters are used for both final models trained with the L1 loss and quantile loss.

\begin{figure}[htbp]
\centering
\includegraphics[width=0.99\columnwidth]{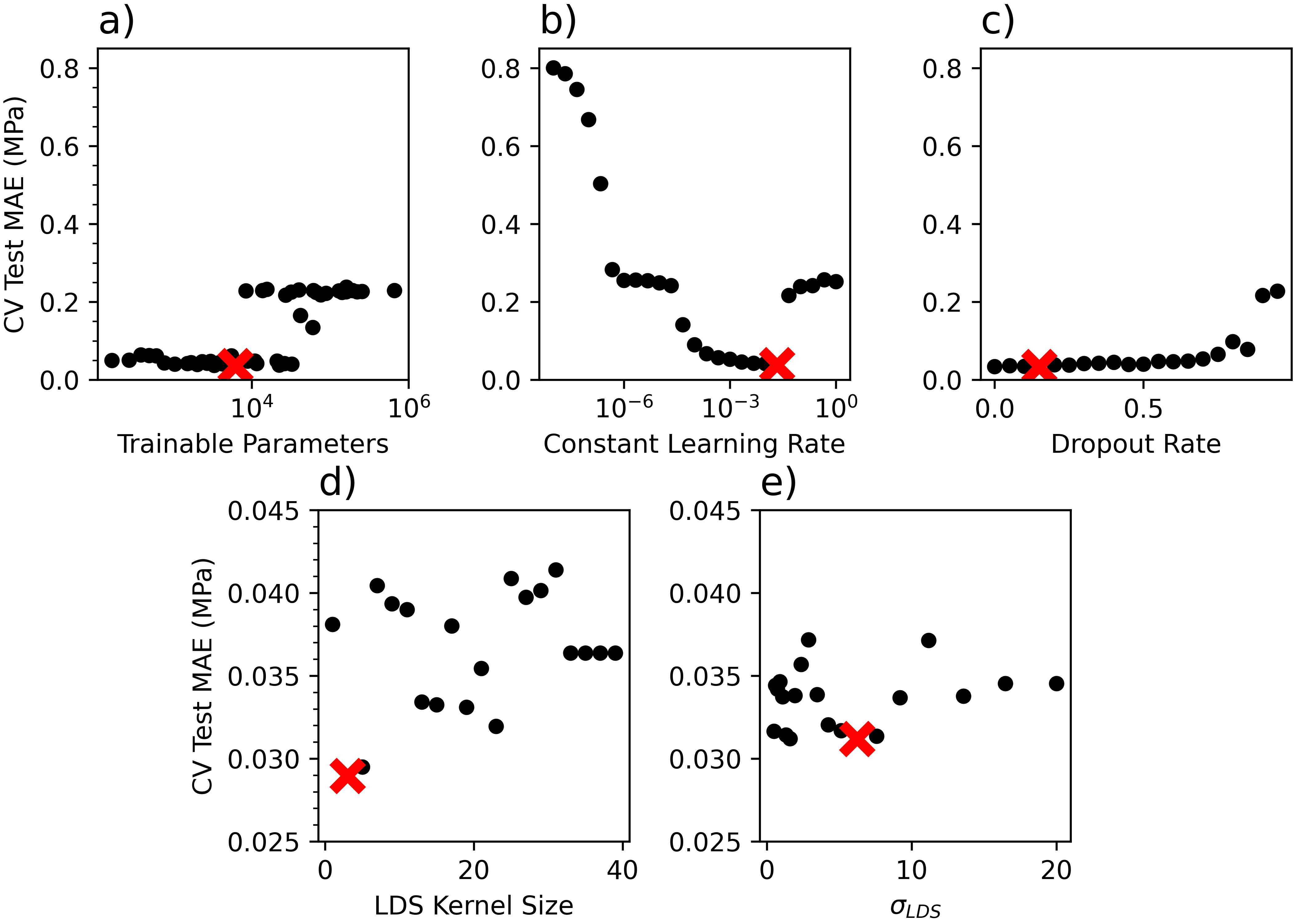}
\caption{Summary of 18-Fold CV MAE loss of the sequential grid search for Jordan data set model hyperparameters examining a) model architecture summarized by number of trainable parameters, b) constant learning rate, c) dropout rate, d) LDS kernel size, and e) $\sigma_{LDS}$. Minimum MAE for each hyperparameter is shown as a larger "X" in red.}
\label{jordan_hyperparams}
\end{figure}

We present results for two separate experiments. The first experiments uses the 18-fold cross-validation method described in Section \ref{approach_and_methods} where the validation and test splits comprised of entire experiments (18 fold due to 18 unique experiments). In other words, an entire temperature or strain-rate is left out of training and the CV errors measure how well the model could interpolate or extrapolate to said value. The results for these models are describe in Table \ref{jordan-our-cv-method-errors-table}.

\begin{table}[htbp]
\caption{Errors for stress functions on Jordan et al. data set using our standard CV partitioning method. MAE, RMSE, Quantile, and SD $\sigma$ have units of MPa and $R^2$ and ACD have units of fractional percentage.} 
\label{jordan-our-cv-method-errors-table}
\begin{center}
\setlength\tabcolsep{5pt}
\begin{tabular}{ccccc}
\hline 
& \multicolumn{2}{c}{\textbf{L1 Loss}} & \multicolumn{2}{c}{\textbf{Quantile Loss}}  \\
& \textbf{18-Fold CV} & \textbf{Ensemble} & \textbf{18-Fold CV} & \textbf{Ensemble} \\
\hline 
\textbf{MAE} & 0.940 & 0.279 & 1.019 & 0.275 \\
\textbf{RMSE} & 1.275 & 0.450 & 1.256 & 0.460 \\
\textbf{$R^2$} & 0.978 & 0.996 & 0.981 & 0.996 \\
\textbf{Quantile} & - & - & 1.340 & 0.449 \\
\textbf{ACD (68\% CI)} & - & - & 0.373 & 0.264 \\
\textbf{ACD (95\% CI)} & - & - & 0.125 & 0.043 \\
\textbf{SD $\sigma$} & 4.437 & 7.095 & 4.437 & 7.095 \\
\hline 
\end{tabular}
\end{center}
\end{table}

However, Jordan et al. choose to randomly partition their train-test split without regard to the experiment of origination. In other words, some data of each strain-rate and temperature are in both the train and test splits during their experiments. In order to compare more directly with their results, we also conduct an 18 fold cross validation where samples are partitioned completely randomly. The results for these models are described in Table \ref{jordan-their-cv-method-errors-table}.

\begin{table}[htbp]
\caption{Errors for stress functions on Jordan et al. data set using using random segmentation for CV. MAE, RMSE, Quantile, and SD $\sigma$ have units of MPa and $R^2$ and ACD have units of fractional percentage.} 
\label{jordan-their-cv-method-errors-table}
\begin{center}
\setlength\tabcolsep{5pt}
\begin{tabular}{ccccc}
\hline 
& \multicolumn{2}{c}{\textbf{L1 Loss}} & \multicolumn{2}{c}{\textbf{Quantile Loss}}  \\
& \textbf{18-Fold CV} & \textbf{Ensemble} & \textbf{18-Fold CV} & \textbf{Ensemble} \\
\hline 
\textbf{MAE} & 0.326 & 0.250 & 0.295 & 0.220 \\
\textbf{RMSE} & 0.526 & 0.419 & 0.477 & 0.379 \\
\textbf{$R^2$} & 0.994 & 0.997 & 0.995 & 0.997 \\
\textbf{Quantile} & - & - & 0.420 & 0.346 \\
\textbf{ACD (68\% CI)} & - & - & 0.149 & 0.232 \\
\textbf{ACD (95\% CI)} & - & - & 0.033 & 0.042 \\
\textbf{SD $\sigma$} & 7.084 & 7.095 & 7.084 & 7.095 \\
\hline 
\end{tabular}
\end{center}
\end{table}

Finally, Jordan et al. examine three polypropylene, uniaxial tension experiments at a strain-rate of $10^{-3} sec^{-1}$ and temperatures of 25$^{\circ}$C, 50$^{\circ}$C and 75$^{\circ}$C to benchmark against Johnson et al. Using $q_{\alpha=0.5}$ from the ensemble model trained based on random partitioning of the data to ensure fair comparison, Figure \ref{Jordan_comp_fig} compares experimental data with our modeled predictions. We find the RMSE to be 0.216. In \cite{JORDAN2020}, the Jordan model had RMSE of 0.29 and the Johnson model had RMSE of 0.78.

\begin{figure}[htbp]
\centering
\includegraphics[width=0.99\columnwidth]{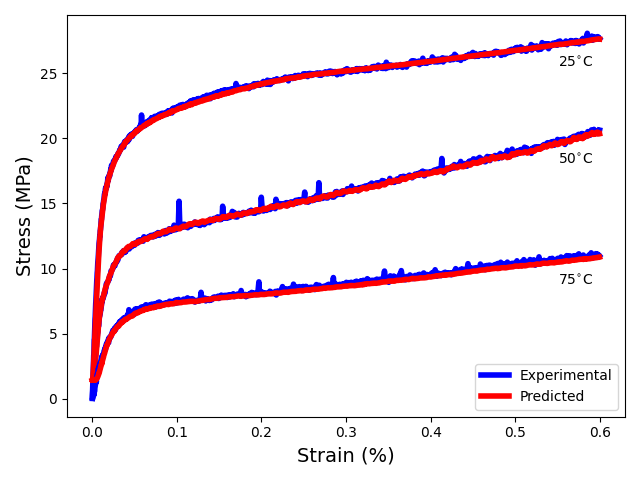}
\caption{Comparing raw experimental and predicted stress-strain curves for Jordan data set at a strain-rate of $10^{-3} sec^{-1}$ and temperatures of 25$^{\circ}$C, 50$^{\circ}$C, and 75$^{\circ}$C.}
\label{Jordan_comp_fig}
\end{figure}

\subsubsection{Benchmarking against Gongoa}

The results of the hyperparameter optimization using the L1 loss function are shown in Figure \ref{BU_lattice_hyperparams}. The chosen hyperparamters are as follows: the model architecture is $(100, 100)$, the learning rate is $4.642 \cdot 10^{-3}$, the dropout rate is $0.25$, the LDS kernel size is $11$, and $\sigma_{LDS}$ is $11.171$. These hyperparameters are used for both final models trained with the L1 loss and quantile loss.

\begin{figure}[htbp]
\centering
\includegraphics[width=0.99\columnwidth]{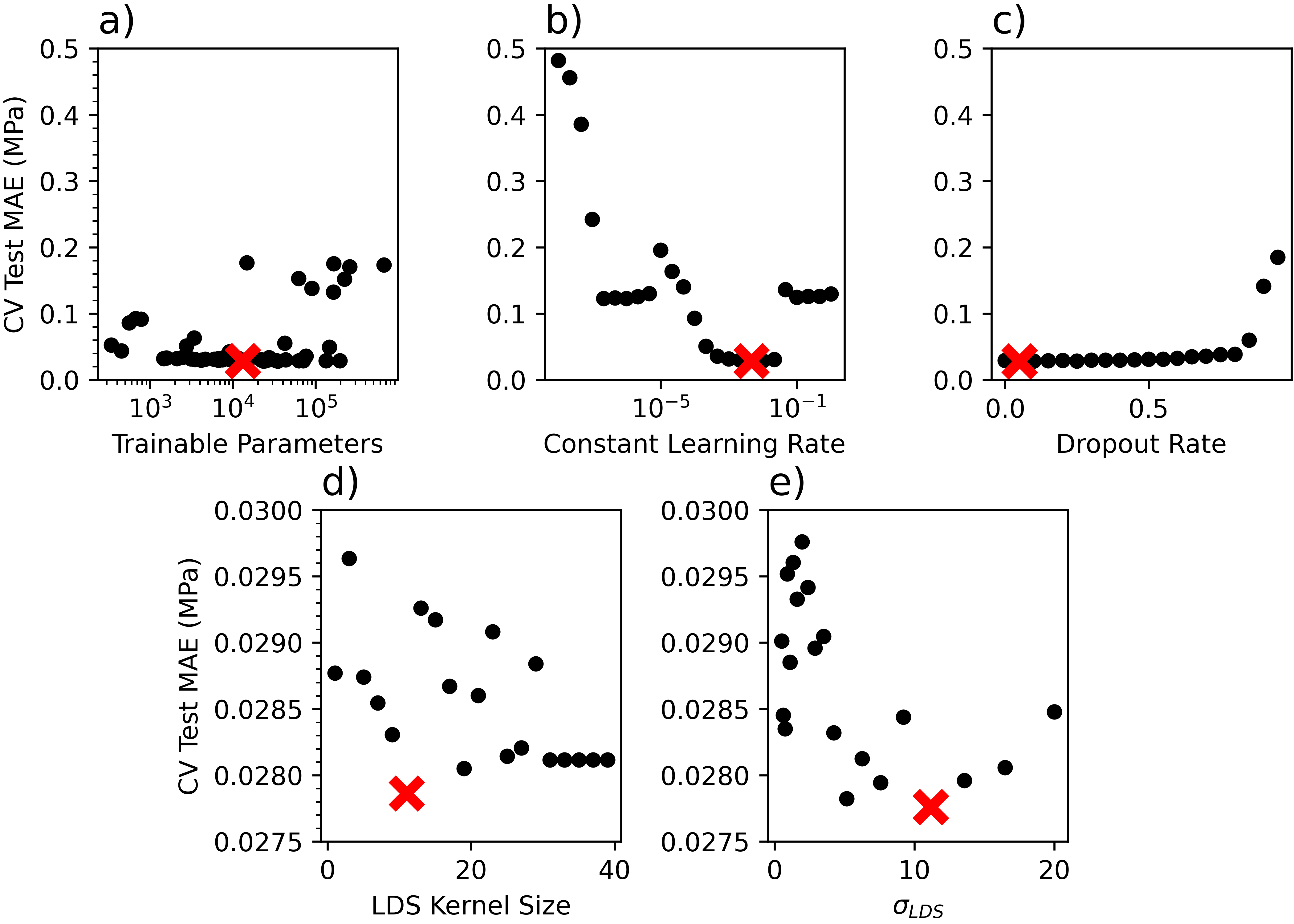}
\caption{Summary of 20-Fold CV MAE loss of the sequential grid search for Gongora data set model hyperparameters examining a) model architecture summarized by number of trainable parameters, b) constant learning rate, c) dropout rate, d) LDS kernel size, and e) $\sigma_{LDS}$. Minimum MAE for each hyperparameter is shown as a larger "X" in red.}
\label{BU_lattice_hyperparams}
\end{figure}

The results for the stress functions are described in Table \ref{gongoa-stress-errors-table} and a parity plot is shown in Figure \ref{BU_stress_parity_plot}. We emphasize that Figure \ref{BU_stress_parity_plot}'s model is trained using the quantile loss and the point estimates presented are from $q_{\alpha=0.5}$. 

\begin{table}[htbp]
\caption{Errors for stress functions on Gongora et al. data set. MAE, RMSE, Quantile, and SD $\sigma$ have units of MPa and $R^2$ and ACD have units of fractional percentage.} 
\label{gongoa-stress-errors-table}
\begin{center}
\setlength\tabcolsep{5pt}
\begin{tabular}{ccccc}
\hline
& \multicolumn{2}{c}{\textbf{L1 Loss}} & \multicolumn{2}{c}{\textbf{Quantile Loss}}  \\
& \textbf{20-Fold CV} & \textbf{Ensemble} & \textbf{20-Fold CV} & \textbf{Ensemble} \\
\hline
\textbf{MAE} & 0.486 & 0.341 & 0.486 & 0.347 \\
\textbf{RMSE} & 0.731 & 0.558 & 0.774 & 0.555 \\
\textbf{$R^2$} & 0.933 & 0.967 & 0.939 & 0.967 \\
\textbf{Quantile} & - & - & 0.647 & 0.459 \\
\textbf{ACD (68\% CI)} & - & - & 0.082 & 0.178 \\
\textbf{ACD (95\% CI)} & - & - & 0.036 & 0.041 \\
\textbf{SD $\sigma$} & 3.058 & 3.065 & 3.058 & 3.065 \\
\hline 
\end{tabular}
\end{center}
\end{table}

\begin{table}[htbp]
\caption{Errors for peak acceleration on Gongora et al. data set. MAE, RMSE, Quantile, and SD Peak Acc. have units of G, RMSPE has units of percent, and $R^2$ and ACD have units of fractional percentage.} 
\label{gongoa-peakG-errors-table}
\begin{center}
\setlength\tabcolsep{5pt}
\begin{tabular}{ccccc}
\hline 
& \multicolumn{2}{c}{\textbf{L1 Loss}} & \multicolumn{2}{c}{\textbf{Quantile Loss}}  \\
& \textbf{20-Fold CV} & \textbf{Ensemble} & \textbf{20-Fold CV} & \textbf{Ensemble} \\
\hline 
\textbf{MAE} & 53.5 & 43.1 & 71.3 & 50.6 \\
\textbf{RMSE} & 70.5 & 56.7 & 84.5 & 67.0 \\
\textbf{RMSPE} & 16.7 & 12.5 & 20.5 & 16.7 \\
\textbf{$R^2$} & 0.442 & 0.598 & 0.362 & 0.464 \\
\textbf{Quantile} & - & - & 102.0 & 66.1 \\
\textbf{ACD (68\% CI)} & - & - & 0.372 & 0.139 \\
\textbf{ACD (95\% CI)} & - & - & 0.287 & 0.063 \\
\textbf{SD Peak Acc.} & 74.2 & 87.5 & 74.2 & 87.5 \\
\hline 
\end{tabular}
\end{center}
\end{table}

Using the approach described in Section \ref{approach_and_methods}, we calculate $a(t)$ curves for each impact. We present the results for peak accelerations in Table \ref{gongoa-peakG-errors-table}. We also examine each impact's $a(t)$ curve using the ensemble quantile model independently and averaged across impacts to calculate errors. We find $R^2 = 0.972$, quantile loss to be 320.7, ACD 68\% and 95\% CI to be 0.222 and 0.048, respectively. 

\begin{figure}[htbp]
% \centering
\includegraphics[width=0.99\columnwidth]{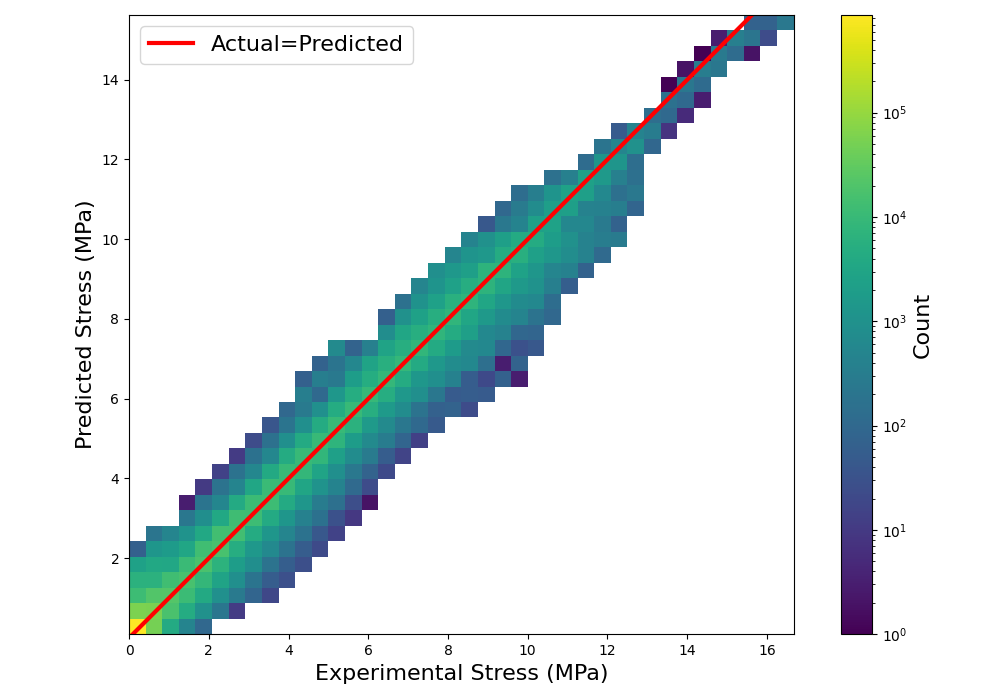}
\caption{Parity plot for stress for Gongora data set based on the quantile loss model where point estimates are $q_{\alpha = 0.5}$ with 40x40 bins.}
\label{BU_stress_parity_plot}
\end{figure}

Finally, Gongoa et al. examine five lattice families that are not included in any training to estimate true validation error. They report RMSPE to be $18\%$. Using our methodology, we find the RMSPE to be $17.7\%$.

\subsubsection{98A Data}

The results of the hyperparameter optimization using the L1 loss function are shown in Figure \ref{98A_hyperparams}. The chosen hyperparamters are as follows: the model architecture is $(200, 100)$, the learning rate is $2.154 \cdot 10^{-3}$, the dropout rate is $0.05$, the LDS kernel size is $17$, and $\sigma_{LDS}$ is $0.50$. These hyperparameters are used for both final models trained with the L1 loss and quantile loss with the exception of two folds using quantile loss where a dropout rate of $0.50$ is used due to significant over fitting. 

\begin{figure}[htbp]
\centering
\includegraphics[width=0.99\columnwidth]{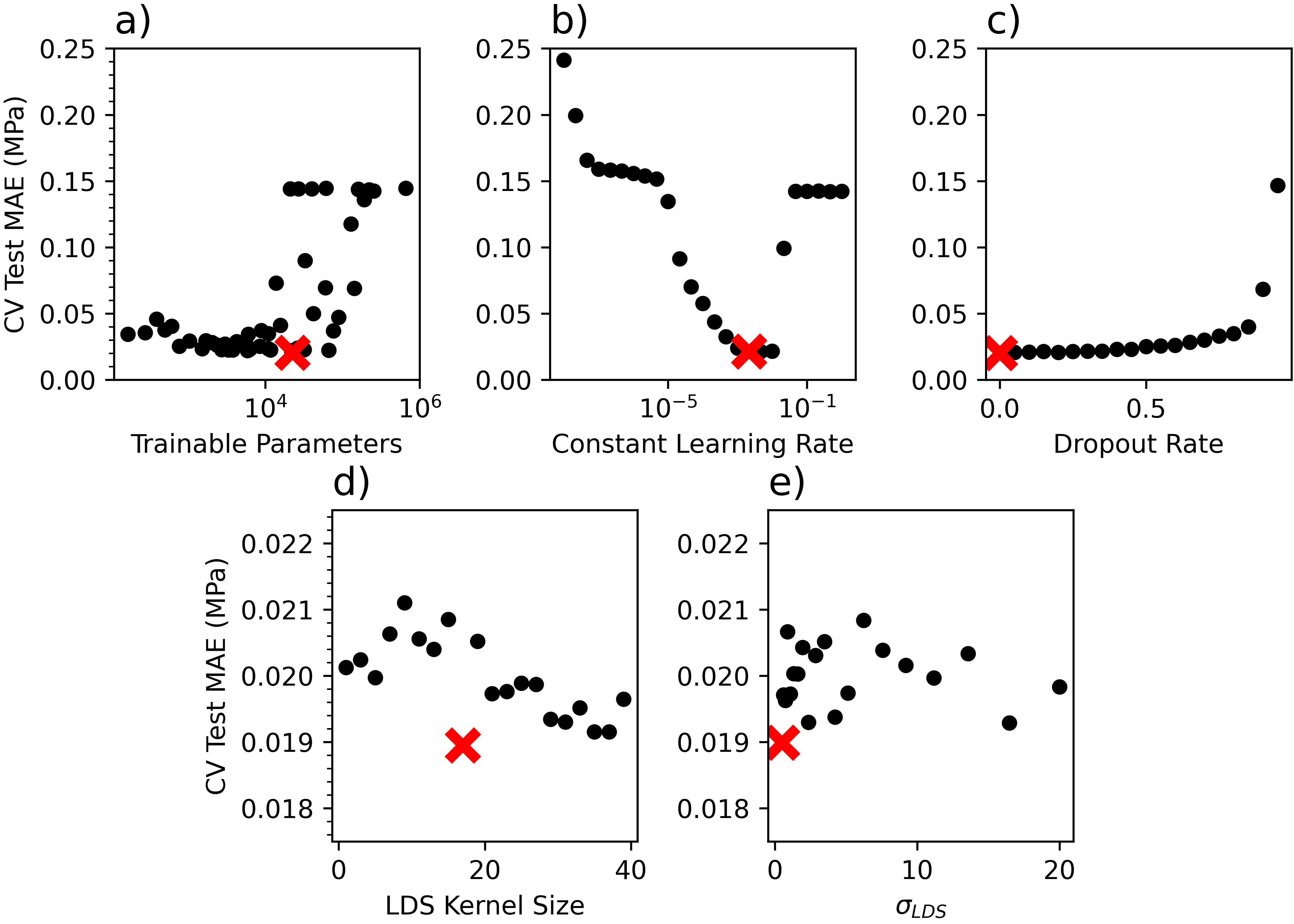}
\caption{Summary of 20-Fold CV MAE loss of the sequential grid search for 98A data set model hyperparameters examining a) model architecture summarized by number of trainable parameters, b) constant learning rate, c) dropout rate, d) LDS kernel size, and e) $\sigma_{LDS}$. Minimum MAE for each hyperparameter is shown as a larger "X" in red.}
\label{98A_hyperparams}
\end{figure}

The results for the stress functions are described in Table \ref{98A-stress-errors-table} and a parity plot is shown in Figure \ref{98A_stress_parity_plot}. We emphasize that Figure \ref{98A_stress_parity_plot}'s model is trained using quantile loss and the point estimates presented are from $q_{\alpha=0.5}$. 

\begin{table}[htbp]
\caption{Errors for stress functions on 98A data set. MAE, RMSE, Quantile, and SD $\sigma$ have units of MPa and $R^2$ and ACD have units of fractional percentage.} 
\label{98A-stress-errors-table}
\begin{center}
\setlength\tabcolsep{5pt}
\begin{tabular}{ccccc}
\hline 
& \multicolumn{2}{c}{\textbf{L1 Loss}} & \multicolumn{2}{c}{\textbf{Quantile Loss}}  \\
& \textbf{20-Fold CV} & \textbf{Ensemble} & \textbf{20-Fold CV} & \textbf{Ensemble} \\
\hline 
\textbf{MAE} & 0.076 & 0.051 & 0.072 & 0.051 \\
\textbf{RMSE} & 0.131 & 0.080 & 0.123 & 0.079 \\
\textbf{$R^2$} & 0.962 & 0.984 & 0.975 & 0.984 \\
\textbf{Quantile} & - & - & 0.095 & 0.069 \\
\textbf{ACD (68\% CI)} & - & - & 0.104 & 0.124 \\
\textbf{ACD (95\% CI)} & - & - & 0.032 & 0.033 \\
\textbf{SD $\sigma$} & 0.689 & 0.630 & 0.689 & 0.630 \\
\hline 
\end{tabular}
\end{center}
\end{table}

We present the results for peak accelerations in Table \ref{98A-peakG-errors-table}. We also examine each impact's $a(t)$ curve using the ensemble the quantile model independently and averaged across impacts to calculate errors. We find $R^2 = 0.979$, quantile loss to be 194.8, ACD 68\% and 95\% CI to be 0.206 and 0.039, respectively.

\begin{figure}[htbp]
% \centering
\includegraphics[width=0.99\columnwidth]{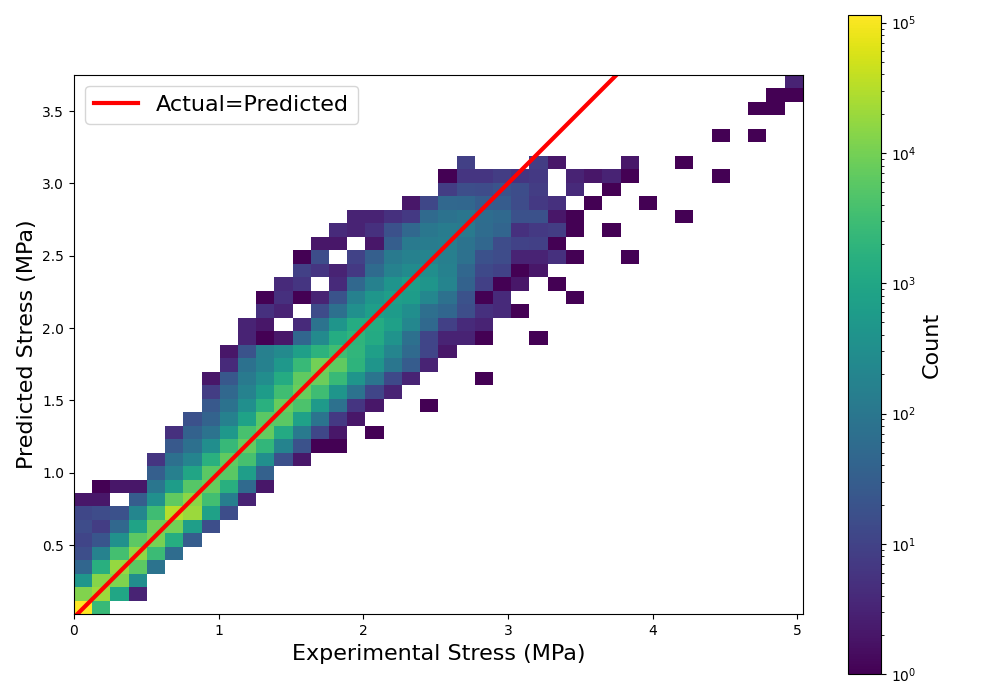}
\caption{Parity plot for stress for 98A data set based on the quantile loss model where point estimates are $q_{\alpha = 0.5}$ with 40x40 bins.}
\label{98A_stress_parity_plot}
\end{figure}

Finally, we map the material parameter space of 98A as a multiobjective optimization according to the loss function in (\ref{loss_func}). The results are shown in Figure \ref{98A_optimizaiton_plot}.

\begin{table}[htbp]
\caption{Errors for peak acceleration on 98A data set. MAE, RMSE, Quantile, and SD Peak Acc. have units of G, RMSPE has units of percent, and $R^2$ and ACD have units of fractional percentage.} 
\label{98A-peakG-errors-table}
\begin{center}
\setlength\tabcolsep{5pt}
\begin{tabular}{ccccc}
\hline 
& \multicolumn{2}{c}{\textbf{L1 Loss}} & \multicolumn{2}{c}{\textbf{Quantile Loss}}  \\
& \textbf{20-Fold CV} & \textbf{Ensemble} & \textbf{20-Fold CV} & \textbf{Ensemble} \\
\hline 
\textbf{MAE} & 31.0 & 15.6 & 27.2 & 16.3 \\
\textbf{RMSE} & 40.6 & 20.5 & 33.8 & 21.1 \\
\textbf{RMSPE} & 13.1 & 7.6 & 10.8 & 7.8 \\
\textbf{$R^2$} & 0.622 & 0.822 & 0.624 & 0.827 \\
\textbf{Quantile} & - & - & 34.5 & 21.1 \\
\textbf{ACD (68\% CI)} & - & - & 0.160 & 0.156 \\
\textbf{ACD (95\% CI)} & - & - & 0.050 & 0.036 \\
\textbf{SD Peak Acc.} & 41.7 & 45.9 & 41.7 & 45.9 \\
\hline
\end{tabular}
\end{center}
\end{table}

\begin{figure}[htbp]
% \centering
\includegraphics[width=0.99\columnwidth]{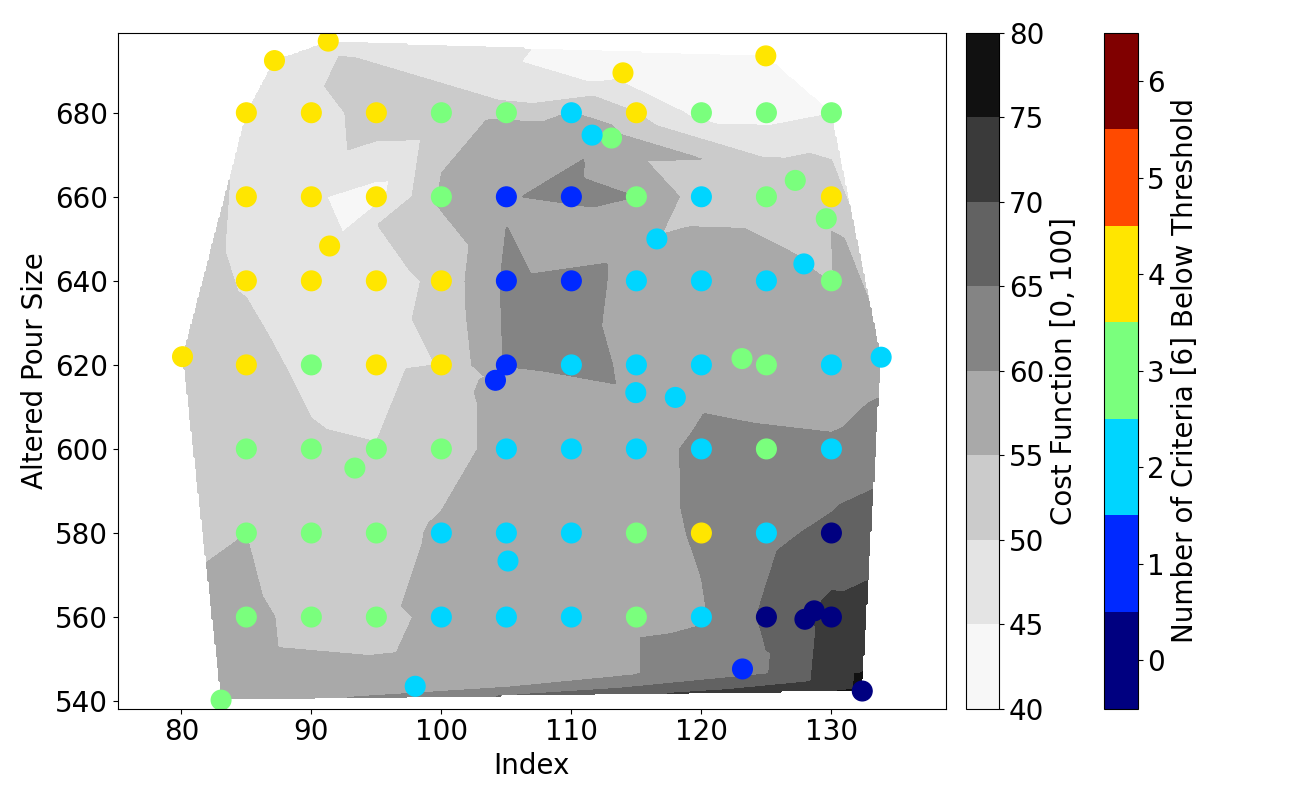}
\caption{Mapping of the 98A material parameter space as a multi-objective optimization where the grey-scale gradients are the loss value, the points represented where samples are calculated, and their color represents how many of the 6 impacts are below their respective threshold.}
\label{98A_optimizaiton_plot}
\end{figure}

\section{Discussion}\label{discussion}

\subsection{Jordan Data Set}

We begin with the Jordan data set comparing our method with the Johnson constitutive model and the Jordan neural network approach. Regardless of the CV partitioning method, we note that the errors for models trained with the L1 vs quantile loss are similar leading us to conclude that it is possible to model variance in the observed data without loss of accuracy for point estimation. 

Most notably, however, is the comparison between our approach shown in Figure \ref{Jordan_comp_fig}. We find the RMSE to be 0.216 noting that it is reported that the Jordan model has RMSE of 0.29 and the Johnson model has RMSE of 0.78 \cite{JORDAN2020}. In other words, we achieve a noticeably lower error than both a leading machine learning and a leading constitute theory based approach. This marked improvement not only helps to validate our methodology, but shows it can achieve lower errors than leading existing methodologies. 

\subsection{Gongoa Data Set}

First and most importantly, we note that our methodology achieves comparably low errors to Gongora et al.'s approach as evidenced in part by our $17.7\%$ true validation error compared with their $18\%$. We emphasize that the authors claim this magnitude of error to be comparable ``with the inherent variation in the fabrication process." 

Our deep learning approach offers significant improvement with regards to the generality of the model's use. As mentioned above, Gongora's model is only capable of predicting impact related criteria for samples at a single impact velocity, drop mass, thickness, and area dictated by their experimental data. Since we model the stress function and employ theory to calculate $a(t)$, we can predict for any reasonable value of the aforementioned parameters; they are independent inputs in our approach. Additionally, we predict the entire acceleration curve, rather than simply its peak value.

We do not require an experimental stress-strain curve as an input unlike the Gongora model. In turn, this removes an experimental requirement should one wish to search the parameter space or complete an optimization based on the model. Further, we achieve these improvements while also capturing the observed variance in the experimental data through quantile regression, another benefit. 

However, we acknowledge that our improvements come at the cost of time, energy, and the need for GPU hardware to train the models as deep neural networks are computationally more expensive than linear regression.

\subsection{98A Data Set}

We present the 98A data set as it is a great case study to examine our proposed methodology. We need to understand the effect of large deformation strain across a large range of strain-rates as well as the effect of temperature and material parameters on stress. Multiple samples are collected at identical impact conditions and over a large range of impact velocities. Finally, the foam is highly variable in a proposed scenario (helmets) where such variation must be taken into account. 

Similar to the above data sets, we note low point estimation errors for the stress function (L1 $< 0.08$ compared with SD $\sigma = 0.69$, $R^2 > 0.96$) for both modeling approaches. When quantile loss is employed in training, we capture the observed variance in data (quantile loss $< 0.095$) without sacrificing point estimation performance. 

When we extend this model to impact mechanics, we note similarly impressive results: L1 $< 31$G compared with SD $\sigma = 46$, $RMSPE < 13\%$ or $7.8\%$ for ensemble models. For the full $a(t)$ curves, we further emphasize an average $R^2 = 0.979$.

Finally, we emphasize the practicality of this model by optimizing the material parameters of a sample of foam under six competing impacts across two impact velocities and three temperatures based on helmet standards \cite{IHSP_spec}. This approach accounts for the uncertainty in the peak acceleration estimate. We note good agreement between the loss function value and the number of impacts out of the six under their respective acceptable threshold. We note that more work can and should be done to better understand the effect of the hyperparameters in the loss function.

\subsection{Conclusions}

In summary, we propose and validate a deep neural network approach to model stress as a state function accounting for observed variance using quantile regression. We note that this approach is not a replacement for classical constitutive theory but offers improvements particularly under the following circumstances: where a theory based model would be difficult to derive and validate, where large data sets can feasibly be collected, where complicated models can reasonably be trained using GPU hardware, or where modeling observed variance is imperative. Future work should look to further validate the methodology on a more diverse array of material types to quantify if deep neural networks can be used for general stress models across domains. Additionally, extensions beyond uniaxial stress should be examined particularly due to their use in finite element modeling.

\section{Acknowledgments}
While there is no direct conflict of interest pertaining to this study, Team Wendy, LLC, has financial interest in head protection products and may incorporate findings of this study into future products.

% \nocite{*}
\bibliographystyle{unsrt}
\bibliography{refs}

\begin{thebibliography}{10}

\bibitem{mechanics-of-materials}
Arthur~P. Boresi and Richard~J Schmidt.
\newblock {\em Advanced Mechanics of Materials}.
\newblock John Wiley \& Sons Inc., 2003.

\bibitem{ghaboussi}
J.~Ghaboussi, J.~H.~Garrett Jr., and X.~Wu.
\newblock Knowledge‐based modeling of material behavior with neural networks.
\newblock {\em Journal of Engineering Mechanics, Vol. 117, No. 1}, 1991.

\bibitem{ALHAIK2006}
M.S. Al-Haik, M.Y. Hussaini, and H.~Garmestani.
\newblock Prediction of nonlinear viscoelastic behavior of polymeric composites using an artificial neural network.
\newblock {\em International Journal of Plasticity}, 22(7), 2006.

\bibitem{Jeremić}
Boris Jeremić and Kallol Sett.
\newblock {\em The Influence of Uncertain Material Parameters on Stress-Strain Response}, pages 132--147.
\newblock ASCE, 2012.

\bibitem{impact-kin}
Steven Rowson, Stefan~M. Duma, Jonathan~G. Beckwith, Jeffrey~J. Chu, Richard~M. Greenwald, Joseph~J. Crisco, P.~Gunnar Brolinson, Ann-Christine Duhaime, Thomas~W. McAllister, and Arthur~C. Maerlender.
\newblock Rotational head kinematics in football impacts: An injury risk function for concussion.
\newblock {\em Annals of Biomedical Engineering}, 40(1):1--13, 2012.

\bibitem{haward-and-thackray}
R.~N. Haward and G.~Thackray.
\newblock The use of a mathematical model to describe isothermal stress-strain curves in glassy thermoplastics.
\newblock {\em Proceedings of the Royal Society of London. Series a, Mathematical and Physical Sciences, Vol. 302, No. 1471}, 1968.

\bibitem{EDWARDS1986}
S.F. Edwards and Th. Vilgis.
\newblock The effect of entanglements in rubber elasticity.
\newblock {\em Polymer}, 27(4):483--492, 1986.

\bibitem{LingWuMei}
Shengbo Ling, Zhen Wu, and Jie Mei.
\newblock Comparison and review of classical and machine learning-based constitutive models for polymers used in aeronautical thermoplastic composites.
\newblock {\em Reviews on Advanced Materials Science}, 62(1):20230107, 2023.

\bibitem{johnson-cook}
G.~R. Johnson and W.~H. Cook.
\newblock A constitutive model and data for metals subjected to large strains, high strain rates and high temperatures.
\newblock {\em Engineering Fracture Mechanics, Vol. 21}, 1983.

\bibitem{JOHNSEN2019}
Joakim Johnsen, Arild~Holm Clausen, Frode Grytten, Ahmed Benallal, and Odd~Sture Hopperstad.
\newblock A thermo-elasto-viscoplastic constitutive model for polymers.
\newblock {\em Journal of the Mechanics and Physics of Solids}, 124:681--701, 2019.

\bibitem{STOFFEL2020}
Marcus Stoffel, Franz Bamer, and Bernd Markert.
\newblock Deep convolutional neural networks in structural dynamics under consideration of viscoplastic material behaviour.
\newblock {\em Mechanics Research Communications}, 108:103565, 2020.

\bibitem{GORJI2020}
Maysam~B. Gorji, Mojtaba Mozaffar, Julian~N. Heidenreich, Jian Cao, and Dirk Mohr.
\newblock On the potential of recurrent neural networks for modeling path dependent plasticity.
\newblock {\em Journal of the Mechanics and Physics of Solids}, 143:103972, 2020.

\bibitem{LI2019320}
Xueyang Li, Christian~C. Roth, and Dirk Mohr.
\newblock Machine-learning based temperature- and rate-dependent plasticity model: Application to analysis of fracture experiments on dp steel.
\newblock {\em International Journal of Plasticity}, 118:320--344, 2019.

\bibitem{JORDAN2020}
Benoit Jordan, Maysam~B. Gorji, and Dirk Mohr.
\newblock Neural network model describing the temperature- and rate-dependent stress-strain response of polypropylene.
\newblock {\em International Journal of Plasticity}, 135:102811, 2020.

\bibitem{GONGORA2022}
Aldair~E. Gongora, Kelsey~L. Snapp, Richard Pang, Thomas~M. Tiano, Kristofer~G. Reyes, Emily Whiting, Timothy~J. Lawton, Elise~F. Morgan, and Keith~A. Brown.
\newblock Designing lattices for impact protection using transfer learning.
\newblock {\em Matter}, 5(9):2829--2846, 2022.

\bibitem{Ivan}
Ivan Hlav\'{a}cek.
\newblock Reliable solution of a perfect plastic problem with uncertain stress-strain law and yield function.
\newblock {\em SIAM Journal on Numerical Analysis}, 39(5):1539--1555, 2002.

\bibitem{BREWICK2018}
Patrick~T. Brewick and Kirubel Teferra.
\newblock Uncertainty quantification for constitutive model calibration of brain tissue.
\newblock {\em Journal of the Mechanical Behavior of Biomedical Materials}, 85:237--255, 2018.

\bibitem{Brizard}
D.~Brizard, S.~Ronel, and E.~Jacquelin.
\newblock Estimating measurement uncertainty on stress-strain curves from shpb.
\newblock {\em Experimental Mechanics}, 57(5):735--742, 2017.

\bibitem{Romero}
Vicente~J. Romero, Benjamin~B. Schroeder, James~F. Dempsey, Nicole~L. Breivik, George~E. Orient, Bonnie~R. Antoun, John~R. Lewis, and Justin~G. Winokur.
\newblock Simple effective conservative treatment of uncertainty from sparse samples of random variables and functions.
\newblock {\em ASME Journal of Risk Uncertainty Part B.}, 2018.

\bibitem{yang2021delving}
Yuzhe Yang, Kaiwen Zha, Ying-Cong Chen, Hao Wang, and Dina Katabi.
\newblock Delving into deep imbalanced regression.
\newblock In {\em International Conference on Machine Learning}, 2021.

\bibitem{Brown_Drop_Tower}
Jialiang Tao.
\newblock Development of a drop tower impact system for characterizing compressible foams under dynamic loads.
\newblock {\em Mechanics of Solids Theses and Dissertations, Brown Digital Repository.}, 2018.

\bibitem{ISLAM2020}
M.A. Islam, M.A. Kader, P.J. Hazell, J.P. Escobedo, A.D. Brown, and M.~Saadatfar.
\newblock Effects of impactor shape on the deformation and energy absorption of closed cell aluminium foams under low velocity impact.
\newblock {\em Materials \& Design}, 191:108599, 2020.

\bibitem{Mines_Charecterization}
D.~Morrison, J.~Morton, M.~Foster, and L.~Lamberson.
\newblock Temperature dependent dynamic response of open-cell polyurethane foams.
\newblock {\em Unpublished}, 2023.

\bibitem{SEA_standard}
SAE International.
\newblock Instrumentation for impact test part 1 - electronic instrumentation.
\newblock Technical report, SAE Standard J211/1\_202208, 2022.

\bibitem{2022arXiv221102989J}
Aryan {Jadon}, Avinash {Patil}, and Shruti {Jadon}.
\newblock {A Comprehensive Survey of Regression Based Loss Functions for Time Series Forecasting}.
\newblock {\em arXiv e-prints}, page arXiv:2211.02989, November 2022.

\bibitem{MAKRIDAKIS2020}
Spyros Makridakis, Evangelos Spiliotis, and Vassilios Assimakopoulos.
\newblock The m4 competition: 100,000 time series and 61 forecasting methods.
\newblock {\em International Journal of Forecasting}, 36(1):54--74, 2020.
\newblock M4 Competition.

\bibitem{IHSP_spec}
US~Army~PM SPE.
\newblock Integrated head protection system ({I}{H}{P}{S}) helmet.
\newblock Technical report, US Army, 2018.

\bibitem{ESTRADA2021}
Jonathan~B. Estrada, Harry~C. Cramer, Mark~T. Scimone, Selda Buyukozturk, and Christian Franck.
\newblock Neural cell injury pathology due to high-rate mechanical loading.
\newblock {\em Brain Multiphysics}, 2:100034, 2021.

\bibitem{background-metrics}
Xianghao Zhan, Yiheng Li, Yuzhe Liu, August~G. Domel, Hossein~Vahid Alizadeh, Zhou Zhou, Nicholas~J. Cecchi, Samuel~J. Raymond, Stephen Tiernan, Jesse Ruan, Saeed Barbat, Olivier Gevaert, Michael~M. Zeineh, Gerald~A. Grant, and David~B. Camarillo.
\newblock Predictive factors of kinematics in traumatic brain injury from head impacts based on statistical interpretation.
\newblock {\em Annals of Biomedical Engineering}, 49(10):2901--2913, 2021.

\end{thebibliography}

\end{document}